\documentclass{article}
\usepackage{graphicx}
\begin{document}

\markboth{E. J. Calegari and S. G. Magalhaes}
{Spectral function of a $d-p$ Hubbard Model}

%
%

\title{SPECTRAL FUNCTION OF A $d-p$ HUBBARD MODEL}

\author{E. J. CALEGARI and S. G. MAGALHAES\footnote{ggarcia@ccne.ufsm.br}\\
{\it{Laborat\'orio de Mec\^anica Estat\'{\i}stica e Teoria da Mat\'eria Condensada}}
\\{\it{Departamento de F\'{\i}sica - Universidade Federal de Santa Maria}}\\
 {\it{97105-900 - Santa Maria, RS, Brazil}}}



\maketitle


\begin{abstract}
This work investigates a d-p Hubbard model by the n-pole approximation in the hole-doped regime. In particular,  the spectral function $A(\omega,\vec{k})$ 
is analyzed varying the filling, the local Coulomb interaction and the $d-p$ hybridization. It should be remarked that the original n-pole approximation (Phys. Rev. 184 (1969) 451) has been improved in order to include adequately the $\vec{k}$-dependence of the important correlation function $\langle \vec{S}_j\cdot\vec{S}_i\rangle$ present in the poles of the Green's functions. It has been verified that the topology of the Fermi surface (defined by $A(\omega=0,\vec{k})$) is deeply affected by the doping, the strength of the Coulomb interaction and also by the hybridization. Particularly, in the underdoped regime, the spectral function $A(\omega=0,\vec{k})$ presents very low intensity close to the anti-nodal points $(0,\pm \pi)$ and $(\pm \pi,0)$. Such a behavior produces an anomalous Fermi surface (pockets) with pseudogaps in the region of the anti-nodal points. On the other hand, if the $d-p$ hybridization is enhanced sufficiently, such pseudogaps vanish. It is precisely the correlation function $\langle \vec{S}_j\cdot\vec{S}_i\rangle$ present in the poles of the Green's functions which plays the important role in the underdoped situation.
In fact, antiferromagnetic correlations coming from $\langle \vec{S}_j\cdot\vec{S}_i\rangle$ strongly modify the quasi-particle band structure.  This is the ultimate source of anomalies in the Fermi surface in the present approach.
\end{abstract}


\section{Introduction}

More than two decades after the discovery of the cuprates \cite{bednorz86}, 
the theoretical description of this phenomenon still represents a challenge for the physicists. 
The fundamental aspects of the cuprates phase diagram can be described in the following terms.
In the overdoped regime, there is a conventional normal metal, while in the very low underdoped, it is displayed an antiferromagnetic insulator (AFI) phase. Between these limits, there is the onset of a superconducting (SC) phase. Nevertheless, it is 
the so-called pseudogap metal \cite{timusk} above the SC and close to the AFI phase which attracts much of the interest. 
It is  widely believed that the understanding of such region is the key element to reveal the nature of the cuprates.

The pseudogap phase is characterized by a density of states with low intensity near the Fermi level leading to an anomalous Fermi surface which is subject of intensive investigations. In the experimental field, while the Fermi surface determined by angle-resolved photoemission spectroscopy (ARPES) is formed by disconnected 'Fermi arcs' \cite{damascelli03}, the quantum-oscillation experiments detect hole-pockets 
enclosing the antinodal points \cite{doiron2007}.
From a theoretical perspective, some general arguments have been proposed to explain the origin of the pseudogap.
The first one relates it to a preformation of electron pairs \cite{Emery95,Randeria08}.
In the second one, the pseudogap phase is originated from a hidden broken symmetry  \cite{Chakravarty02,Varma06}. 
However, there is also a proposition which suggests that  
the presence of short-range antiferromagnetic (AF) correlations 
could be the mechanism responsible for the pseudogap \cite{Kampf90,Morinari09}.
For instance, a recent work \cite{Harrison07} has indicated
that antiferromagnetic fluctuations with short correlation length could play a fundamental role in order to understand 
the apparent disagreement between ARPES and quantum-oscillation experiments.
Therefore, one can set the question how that contributions coming from the  antiferromagnetic correlations, 
could give rise to the pseudogap and consequently, the anomalies on the 
Fermi Surface (FS).

Among the available many body techniques, the n-pole approximation \cite{Roth69,Edwards95,jnunes}
seems to be quite suitable to such a purpose (see also discussion in reference \cite{nolting}). The structure of the poles of the one particle Green's function presents an energy shift (or band shift) in the self energy
composed basically by three terms, one of them is the correlation
function $\langle\vec{S}_{i}\cdot\vec{S}_{j}\rangle$, which can produce AF correlations. 
Actually, recent works have investigated the role of AF correlations using 
the Composite Operator Method (COM)\cite{avella2007,avella09} which is an analytical technique that allows calculating all correlations functions involved in the problem, fully self-consistent. As main result, it is obtained  that AF correlations can be, indeed, a source of a pseudogap. 

It should be stressed that the previous results have been obtained for the one-band Hubbard model.  
One remaining issue is related to the role of hybridization on the AF correlations.
A recent experimental work \cite{chou04}, using the resonant inelastic x-ray scattering (RIXS) technique, has evidenced that  
the hybridization (between the $d$-orbitals of the cooper $Cu$ and the $p$-orbitals of the oxygen $O$) in cuprates systems is stronger than that one expected from tight-binding theory. However, the one-band models neglect the presence of the oxygen sites. Due to the strong correlations at the $Cu$-sites, the oxygen sites may be occupied by holes when the system is doped \cite{ref3}.
Therefore, although the one-band models are able to capture many of the most important physical properties of cuprates, probably,
a model which can incorporate additional details such as the hybridization would be a more adequate description of 
the cuprates. 

In the present work, an extended $d-p$ Hubbard model is treated by a n-pole approximation \cite{Roth69}. Therefore, 
one can investigate how AF correlations 
could affect the properties of the model, mainly focusing on theirs role as a possible  source of a pseudogap and anomalous FS. 
It should be remarked that in the extended $d-p$ Hubbard model used in the present work the model parameters which can affect 
AF correlations are not only the doping $\delta$ and the strength of local Coulomb interaction $U$, but also the hybridization.  
Consequently, it would be possible to produce a systematic investigation of how the variation of these model 
parameters as $\delta$, $U$ and, particularly, the hybridization, can affect
the AF correlations and how such correlations influence the features of 
physical quantities such as the spectral function $A(w,\vec{k})$.

In fact, the extended $d-p$ Hubbard model has been previously studied within
the n-pole approximation \cite{calegari2004,calegari2005EB},  in which $d_{x^2-y^2}$-wave superconductivity has been treated 
following the approach proposed in Ref. \cite{Edwards95}. In Ref. \cite{calegari2005EB}, 
it has been obtained that the increase of hybridization suppresses the superconductivity. This occurs mainly due to the fact that the hybridization has strong effects in the band shift which moves the superconducting gap to low energies breaking the gap symmetry relative to the zero energy. Furthermore, the hybridization broadens and suppresses the density of states becoming unfavored the pair formation (near the Fermi energy).
It is important to mention that in the present work, the $d-p$ model includes hoppings to the second-nearest-neighbors which improves significantly the uncorrelated band structure considered here.

It is well known from earlier results \cite{nolting}  that in the normal state
the function $\langle \vec{S}_j\cdot \vec{S}_i\rangle$ is strongly dependent 
on the doping and the Coulomb interaction, particularly,
on the underdoped regime. Moreover, it has also been verified that the correlation function $\langle S_j^+S_i^-\rangle$, which behaves like $\langle \vec{S}_j\cdot \vec{S}_i\rangle$, is also strongly dependent 
on the $d-p$ hybridization $t^{dp}$ \cite{calegariPB06}. However, for a given doping $\delta$, 
$U$ and $t_{pd}$ produce effects in opposite direction, i.e., while $U$ increases the intensity of $\langle \vec{S}_j\cdot \vec{S}_i\rangle$, 
$t^{dp}$ decreases it. 
The fact that $t^{dp}$ depletes the correlation function $\langle \vec{S}_j\cdot \vec{S}_i\rangle$ and consequently the antiferromagnetic correlations makes the hybridization an adequate physical mechanism to verify whether the AF correlations are or are not essential to the onset of the pseudogap.

 It is important to highlight that, in the present scenario, the AF correlations are deeply influenced by the momentum structure of the spin-spin correlation function  $\langle \vec{S}_j\cdot \vec{S}_i\rangle$. Indeed, there is a considerable number of works treating about this important subject. For instance, in references \cite{AMoreo95,gubernatis01,scalettar09} the behavior of the AF correlations associated to $\langle \vec{S}_j\cdot \vec{S}_i\rangle$ were studied by using Monte Carlo simulations. However, the projection techniques \cite{plakida} and methods like slave-boson \cite{zimmermann} and exact diagonalization \cite{Leung} have also been considered. Actually, quite recent results \cite{avella2007} obtained within the composite operator method (COM), which is an analytical and fully self-consistent method, have shown that the momentum dependence of the charge and spin correlation functions is very important to obtain anomalies like the pseudogap in the underdoped regime of the hole-doped case. Nevertheless, in the original Roth's procedure \cite{Roth69} and in the subsequent works in which the Roth's n-pole approximation is used, it was considered $t_{ij}^{d}=t^{d}$ (where $t_{ij}^{d}$ is the hopping between nearest-neighbors) for the $z$ nearest-neighbors.  Consequently, several correlation functions  present in the band shift (including $\langle \vec{S}_j\cdot \vec{S}_i\rangle$) are constant over all the first Brillouin zone. However, such a simplification is unable to capture properly the $\vec{k}$-dependence of the band shift $W_{\vec{k}\sigma}^d$, mainly, in the underdoped regime. Thus, in the present work, the band shift and, therefore, the $\langle \vec{S}_j\cdot \vec{S}_i\rangle$ have been evaluated including adequately  such $\vec{k}$-dependence.

This paper is structured as follows. Section 2 presents the model and the general
formalism of the Roth's n-pole approximation \cite{Roth69}. In section 3, 
the numerical results are presented and discussed. Finally, section 4 shows a summary and some concluding remarks.

\section{Model and General Formulation}
The Hamiltonian model proposed here \cite{calegari2004,calegari2005EB} is given by:
\begin{equation}
{\cal{H}}=H+H'
\label{eq0}
\end{equation} 
with,
\begin{eqnarray}
H&=&\sum_{i,\sigma }(\varepsilon _{d}-\mu)d_{i\sigma
}^{\dag}d_{i\sigma }+\sum_{\langle i\rangle j,\sigma }t_{ij}^{d}d_{i\sigma
}^{\dag}d_{j\sigma }+
U\sum_{i}n_{i\uparrow}^{d}n_{i\downarrow}^{d}\nonumber\\
 & &+\sum_{i,\sigma }(\varepsilon _{p}-\mu)p_{i\sigma }^{\dag}p_{i\sigma
}+\sum_{\langle i\rangle j,\sigma }t_{ij}^{p}p_{i\sigma }^{\dag}p_{j\sigma }
+\sum_{\langle i\rangle j,\sigma }t_{ij}^{pd}\left( d_{i\sigma
}^{\dag}p_{j\sigma +}p_{i\sigma }^{\dag}d_{j\sigma }\right)
\label{eq1}
\end{eqnarray}
where $\mu$ is the chemical potential. The term $H'$, which represents 
the second-nearest-neighbors, is written as:
\begin{eqnarray}
H'&=&\sum_{\langle\langle i\rangle\rangle j,\sigma }t_{ij}^{\ell d}d_{i\sigma
}^{\dag}d_{j\sigma }+\sum_{\langle\langle i\rangle\rangle j,\sigma }
t_{ij}^{\ell p}p_{i\sigma }^{\dag}p_{j\sigma }.   
\label{eq1.1}
\end{eqnarray}
The $d_{i\sigma}^{\dag}(d_{i\sigma})$ and 
$p_{i\sigma}^{\dag}(p_{i\sigma})$  are the creation(annihilation) operators for {\it electrons}
with spin $\sigma$ in a site $i$. The quantity $U$ stands for the local Coulomb interaction 
between two $d$-electrons with opposite spins. The model (\ref{eq1}) considers a small
$t_{ij}^d$ hopping between 
$d$-orbitals and a large 
$t_{ij}^p$ hopping between
$p$-orbitals. The quantity $t_{ij}^{dp}$ stands for
hopping between $d$- and $p$-orbitals, and $\Delta_{dp}=\varepsilon_p-\varepsilon_d$ represents the on site-energy 
difference between the $d$- and $p$-orbitals,
respectively. The parameters $t_{ij}^{\ell d}$ and $t_{ij}^{\ell p}$ presented in $H'$ represent the hoppings to the second-nearest-neighbors for $d$ and $p$ 
electrons, respectively. The symbols $\langle ...\rangle$ $\left(\langle\langle ...\rangle\rangle\right)$
denote the sum over the first(second)-nearest-neighbors of $i$.
For a rectangular two-dimensional lattice, the $p$-dispersion relation for the first-nearest-neighbors 
is given by:
\begin{equation}
\varepsilon_{1\vec{k}}^p = 2t^p[\cos(k_xa)+\cos(k_ya)]
\label{ep1}
\end{equation}
where $a$ is the lattice parameter.
Considering the homothetic relation\cite{Japiassu92,Caixeiro08} for the $p$ and $d$ dispersion relations, $\varepsilon_{\vec{1k}}^d = \alpha \varepsilon_{1\vec{k}}^p$,
where $\alpha$ is a phenomenological parameter less than the unity. If the hoppings to the second-nearest-neighbors 
are taken into account, 
\begin{equation}
\varepsilon_{\vec{k}}^p = \varepsilon_{1\vec{k}}^p+4t^{\ell p}\cos(k_xa)\cos(k_ya)~~~\mbox{and}~~~ \varepsilon_{\vec{k}}^d = \varepsilon_{\vec{1k}}^d+4t^{\ell d}\cos(k_xa)\cos(k_ya).
\label{ep}
\end{equation}

In order to obtain the Green's functions within the Roth's method \cite{Roth69}, 
it is necessary to define a set of operators $\{A_n\}$ that describes 
adequately the relevant one-particle excitations of the system. 
The set of three operators considered here is $\left\{d_{i\sigma },n_{i-\sigma }^{d}d_{i\sigma },p_{i\sigma }\right\}$.
These operators must satisfy, within some approximations,
the relation  $\left[ A_{n},H\right] _{-}=\sum_{m}K_{nm}A_{m}$, where $A_n$ are 
the operators of the set $\{A_n\}$ defined above. This set of three operators results in a three-poles approximation for the Green's functions,
which in matrix notation is written as:
\begin{equation}
{\bf G}\left( \omega \right) =\widetilde{\bf G}(\omega){\bf N}
\label{eq2.7}
\end{equation}
where
\begin{equation}
\widetilde{\bf G}\left( \omega \right) ={\bf N}(\omega {\bf N}-{\bf E})^{-1}
\label{eq2.8}.
\end{equation}
Here, {\bf E} and {\bf N} are the energy and the normalization matrices given by
\begin{equation}
E_{nm}=\left\langle\left[ \left[ A_{n},H\right] _{-},
A_{m}^{\dag}\right] _{\left( +\right) }\right\rangle~~~~~\mbox{and}~~~~~
N_{nm}=\langle [ A_{n},A_{m}^{\dag}]_{\left( +\right) }\rangle
\label{Enm}
\end{equation}
where $[...,...]_{(+)-}$ denote the (anti)commutator, and $\langle ...\rangle$, the thermal average.   

Considering the set of operators $\{A_n\}$ introduced above and the Hamiltonian given by equation (\ref{eq0}), the energy matrix is:
\begin{eqnarray}
{\bf E}=\left[
\begin{tabular}{ccc}
$\overline{\varepsilon}_{d} +\varepsilon_{\vec{k}}^{d} + Un_{-\sigma}^d$ &
$(\overline{\varepsilon}_{d} + \varepsilon_{\vec{k}}^{d}+U)n_{-\sigma}^d$ & $V_{\vec{k}}^{dp}$\\
\\$(\overline{\varepsilon}_{d} + \varepsilon_{\vec{k}}^{d}+U)n_{-\sigma}^d$ &
$Un_{-\sigma}^d + \Gamma_{\vec{k}-\sigma}$ & $n_{-\sigma}^dV_{\vec{k}}^{dp}$\\ \\
$V_{\vec{k}}^{pd}$ & $n_{-\sigma}^dV_{\vec{k}}^{pd}$ &$\overline{\varepsilon}_{p} +\varepsilon_{\vec{k}}^{p}$
\end{tabular}
\right]
\label{E}
\end{eqnarray}
and the normalization one:
\begin{eqnarray}
{\bf N}=\left[
\begin{tabular}{ccc}
$1$ & $n_{-\sigma}^d$ & $0$\\ \\
$n_{-\sigma}^d$ & $n_{-\sigma}^d$ & $0$\\ \\
$0$ & $0$ &$1$
\end{tabular}
\right]
\label{N}
\end{eqnarray}
with $\overline{\varepsilon}_{d}=\varepsilon_{d}-\mu$. The $V_{\vec{k}}^{dp}(V_{\vec{k}}^{pd})$ are the
Fourier transform of $t_{ij}^{dp}(t_{ij}^{pd})$. It is assumed that the system considered here is translationally invariant,
then $n_{-\sigma}^d=n_{i-\sigma}^d$.
Finally, the quantity $\Gamma_{\vec{k}-\sigma}$ is defined as:
\begin{equation}
\Gamma_{\vec{k}-\sigma}=\overline{\varepsilon}_{d}n_{-\sigma}^d+\varepsilon_{\vec{k}}^{d}(n_{-\sigma}^d)^2+n_{-\sigma}^d(1-n_{-\sigma}^d)W_{\vec{k}\sigma}
\label{gamma}
\end{equation}
where $W_{\vec{k}\sigma}$ is the band shift that will be introduced later on. 

One of the most important elements of the Green's function matrix is the element 
\begin{equation}
G_{\vec{k}\sigma}^{(11)}(\omega)=\frac{(\omega-E_{33})A_{1\vec{k}}(\omega)}{(\omega-E_{33})D_{\vec{k}\sigma}(\omega)-A_{1\vec{k}}(\omega)
V_{\vec{k}}^{pd}V_{\vec{k}}^{dp}}
\label{G11}
\end{equation}
where,
\begin{equation}
A_{1\vec{k}}(\omega)=n_{-\sigma}^d(1-n_{-\sigma}^d)[\omega-\overline{\varepsilon}_{d}-U(1-n_{-\sigma}^d)-W_{\vec{k}\sigma}]
\label{A1}
\end{equation}
and
\begin{eqnarray}
D_{\vec{k}\sigma}(\omega)&=&n_{-\sigma}^d(1-n_{-\sigma}^d)\nonumber\\
& &\times[(\omega-\overline{\varepsilon}_{d}-\varepsilon_{\vec{k}}^{d})
(\omega-\overline{\varepsilon}_{d}-U-W_{\vec{k}\sigma})-Un_{-\sigma}^d(\varepsilon_{\vec{k}}^{d}-W_{\vec{k}\sigma})]
\label{D}.
\end{eqnarray}
The quantity $E_{33}$ is an element of the energy matrix given in equation (\ref{E}).

Considering the Green's function defined in equation (\ref{G11}), the spectral function can be defined as:
\begin{equation}
A_{\sigma}(\vec{k},\omega)=-\frac{1}{\pi}\mbox{Im}[{G_{\vec{k}\sigma }^{(11)}(\omega)}].
\label{Awks}
\end{equation}

In the real space, the band shift presented in equation (\ref{gamma}) is written as:
\begin{equation}
W _{ij-\sigma }=W_{ij-\sigma }^{d}+W_{ij-\sigma }^{pd}
\label{eq4.0}
\end{equation}
where $W_{ij-\sigma }^{pd}$ is given by equation (51) in Ref. \cite{calegari2005EB}.
However, in Ref. \cite{calegari2005EB}, the hopping to second-nearest-neighbors has not been considered, therefore the band shift $W_{\vec{k}\sigma }^d$ presented in equation (\ref{gamma}) must be changed by:
\begin{equation}
n_{\sigma}^d(1-n_{\sigma}^d)W_{\vec{k}\sigma }^d =h_{1\sigma}+\sum_{\langle i=0\rangle j\neq 0}t_{0j}^{d}e^{i\vec{k}\cdot\vec{R}_j}h_{2j\sigma}+\sum_{\langle\langle i=0\rangle\rangle j\neq 0}t_{0j}^{\ell d}e^{i\vec{k}\cdot\vec{R}_j}h_{2j\sigma}
\label{Wd}
\end{equation}
which take into account the hopping to second-nearest-neighbors.
The quantities $h_{1\sigma}$ and $h_{2j\sigma}$ are defined as:
\begin{equation}
h_{1\sigma}= -\sum_{\langle i=0\rangle j\neq 0}t_{0j}^{d}(n_{j0\sigma }^{d} - 2m_{j\sigma })
-\sum_{\langle\langle i=0\rangle\rangle j\neq 0}t_{0j}^{\ell d}(n_{j0\sigma }^{d} - 2m_{j\sigma })
\label{eqh2}
\end{equation}
and
\begin{equation}
h_{2j\sigma}= B_{j\sigma} + \langle \vec{S_j}\cdot\vec{S_0}\rangle 
\label{h2}
\end{equation}
with
\begin{eqnarray}
B_{j\sigma}&=&-\langle S_j^zS_0^z \rangle-\frac{\alpha_{j\sigma}n_{0j\sigma}^d + \beta_{j\sigma}m_{j\sigma}}{1-\beta_{\sigma}\beta_{-\sigma}}  
-\frac{\alpha_{j\sigma}n_{0j-\sigma}^d +\beta_{j\sigma}(n_{0j-\sigma}^d
-m_{j-\sigma} ) }{1-\beta_{\sigma}}.\nonumber\\
\label{Ajs}
\end{eqnarray}

The spin-spin correlation function introduced in equation (\ref{h2}) is given by:
\begin{equation}
\langle \vec{S_j}\cdot\vec{S_0}\rangle= \frac{1}{2}\left(\langle S_j^+S_0^-\rangle 
+\langle S_j^-S_0^+\rangle\right) +\langle S_j^zS_0^z \rangle 
\label{SjSi}.
\end{equation}
Particularly, in the paramagnetic state, $\langle S_j^+S_0^-\rangle=\langle S_j^-S_0^+\rangle$,
and the spin-spin correlation function $\langle \vec{S_j}\cdot\vec{S_0}\rangle$ can be written as:
\begin{equation}
\langle \vec{S_j}\cdot\vec{S_0}\rangle= \langle S_j^+S_0^-\rangle +\langle S_j^zS_0^z \rangle 
\label{SjSi2}
\end{equation}
where,
\begin{equation}
\langle S_j^+S_0^- \rangle =\langle d_{j\sigma}^{\dagger}d_{j-\sigma}d_{0-\sigma}^{\dagger}d_{0\sigma} \rangle=-\frac{\alpha_{j\sigma}n_{0j,-\sigma}^d+\beta_{j,\sigma}m_{j,-\sigma} }{1+\beta_{\sigma}}
\label{S+S-}
\end{equation}
and
\begin{equation}
\langle S_j^zS_0^z \rangle = \frac{(1-\beta_{-\sigma})}{2}\left[(n_{\sigma}^d)^2-\frac{\alpha_{j\sigma}
n_{0j\sigma}+\beta_{j\sigma}m_{j\sigma} ) }{1-\beta_{\sigma}\beta_{-\sigma}}\right]
-\frac{\alpha_{-\sigma}n_{\sigma}^d}{2} 
\label{SzSz}.
\end{equation}

In Refs. \cite{Edwards95,calegari2004,calegari2005EB,calegariPB06}, the band shift has been evaluated following the original Roth's procedure \cite{Roth69}, where  $t_{0j}^{d}=t^{d}$ for the $z$ nearest-neighbors has been considered. Consequently, this procedure removes the $\vec{k}$-dependence of the correlation functions (mainly $\langle \vec{S_j}\cdot\vec{S_i}\rangle$ )  present in the band shift. However, the momentum dependence of such correlation functions  
is a fundamental ingredient to investigate anomalous properties 
like the pseudogap in the underdoped regime of the Hubbard models \cite{avella2007,avella2003}. 
Therefore, in the present work, the band shift defined in equation (\ref{Wd}) is rewritten as:
\begin{eqnarray}
W_{\vec{k}\sigma }^d &=&\frac{1}{n_{\sigma}^d(1-n_{\sigma}^d)}\frac{1}{L}\sum_{\vec{q}}\epsilon(\vec{k}-\vec{q})
F_{\sigma}(\vec{q}),
\label{Wdk}
\end{eqnarray}
where the momentum dependence of the correlation functions has been maintained. The $\epsilon(\vec{k}-\vec{q})$ is given by
\begin{equation}
\epsilon(\vec{k}-\vec{q})=\sum_{\langle i=0\rangle j\neq 0}t_{0j}^de^{i(\vec{k}-\vec{q})\cdot \vec{R}_j}+\sum_{\langle\langle i=0\rangle\rangle j\neq 0}t_{0j}^{\ell d}e^{i(\vec{k}-\vec{q})\cdot \vec{R}_j}
\label{ekq}
\end{equation}
and $F_{\sigma}(\vec{q})$ is given in terms of the Fourier transform of $n_{j0\sigma }^{d}$, $m_{j\sigma}$, $\alpha_{j\sigma}$ and $\beta_{j\sigma}$ introduced in equations (\ref{eqh2})-(\ref{SzSz}) and defined as:
\begin{equation}
n_{0j\sigma}^d=\langle d_{0\sigma}^{\dagger}d_{j\sigma}\rangle=\frac{1}{L}
\sum_{\vec k}{\cal{F}}_{\omega}G_{\vec{k}\sigma}^{(11)}e^{i\vec{k}\cdot \vec{R}_j}
\label{eq16},
\end{equation}
\begin{equation}
m_{j\sigma}=\langle d_{0\sigma}^{\dagger}n_{j-\sigma }^{d}d_{j\sigma}\rangle=\frac{1}{L}
\sum_{\vec k}{\cal{F}}_{\omega}G_{\vec{k}\sigma}^{(12)}e^{i\vec{k}\cdot \vec{R}_j}
\label{eq17},
\end{equation}
\begin{equation}
\alpha_{j\sigma}=\frac{1}{L}
\sum_{\vec k}{\cal{F}}_{\omega}\widetilde{G}_{\vec{k}\sigma}^{(11)}e^{i\vec{k}\cdot \vec{R}_j}
\label{eq18}
\end{equation}
and
\begin{equation}
\beta_{j\sigma}=\frac{1}{L}
\sum_{\vec k}{\cal{F}}_{\omega}\widetilde{G}_{\vec{k}\sigma}^{(12)}e^{i\vec{k}\cdot \vec{R}_j}
\label{eq19}
\end{equation}
where ${\cal{F}}_{\omega}\Gamma(\omega)\equiv\frac{1}{2\pi i}\oint d\omega f(\omega)\Gamma(\omega)$, in which $f(\omega)$ is the Fermi function and $\Gamma(\omega)$ a general Green's function. The Green's functions $G$ and $\widetilde{G}$ are obtained from the definitions (\ref{eq2.7}) and (\ref{eq2.8}), respectively. $L$ represents the number of lattice sites. 

\section{\label{sec:Numresults} Numerical Results}
In this section, a detailed investigation of the Fermi surface associated with the spectral functions for hole-doped regime is done.
As a starting point, the $d-p$ hybridization has been considered $\vec{k}$-independent \cite{calegari2005EB}
$(2V_0^{pd})^2=\langle V_{\vec{k}}^{dp}V_{\vec{k}}^{pd}\rangle$. Here, 
$\langle ...\rangle$ is the average over the first Brillouin zone.
The remaining model parameters are within reasonable ranges estimated for cuprates \cite{huang2001}.
In particular, the parameters
$a=1$, $\varepsilon_d=0$, $\varepsilon_p=3.6$ eV, $t^p=-0.7$ eV, $\alpha= 0.715$ and $t^{\ell p}=0$ have been kept the same for all results presented here. The homothetic dispersion relation considered in the present work (see equations (\ref{ep1}) and (\ref{ep})), indeed, signify that $t^d=\alpha t^p$, therefore $t^d\simeq-0.5$ eV.

\begin{figure}
 \centering
\includegraphics[angle=-90,width=12 cm]{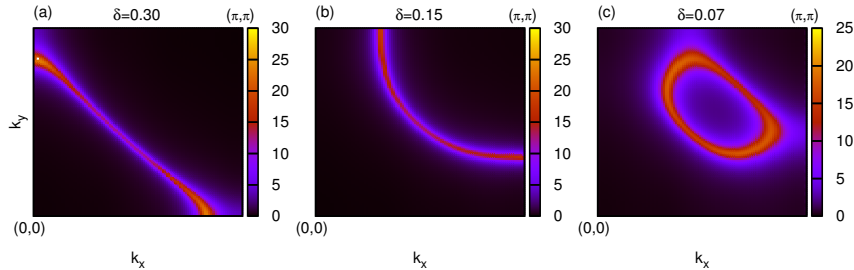}
\caption{\label{fig1}The spectral function $A(\vec{k},\omega = 0)$ representing the Fermi surface for different dopings $\delta$. The model parameters considered here are $U=10$ eV, $V_0^{pd}=1.0$ eV, $t^d=-0.5$ eV, $t^{\ell d}=0.1$ eV and $k_BT=0.01$ eV ($k_{B}$ is the Boltzmann constant).}
\end{figure}

Figure \ref{fig1} shows the Fermi surface for three different doping levels $\delta$, where
$\delta=1-n_T$ (with $n_T=n^d_{\sigma}+n^d_{-\sigma}$). In figure 
\ref{fig1}(a), $\delta=0.30$, and a well defined electron-like Fermi surface is observed. In figure \ref{fig1}(b), $\delta=0.15$
and the nature of the Fermi surface changes to hole-like. However, it is in the underdoped regime,  $\delta=0.07$, that the topology of the Fermi surface changes drastically with the emergence of a hole-pocket enclosing the nodal point $(\frac{\pi}{2},\frac{\pi}{2})$. As a consequence, due to low spectral intensity, a pseudogap emerges near the antinodal points $(\pi,0)$ and $(0,\pi)$, as shown in figure \ref{fig1}(c).
This result can be better understood by analyzing the features of the lower quasiparticle band which is strong affected by the antiferromagnetic correlations associated with the correlation function $\langle \vec{S_j}\cdot\vec{S_i}\rangle$.
\begin{figure}
 \centering
\includegraphics[angle=-90,width=13 cm]{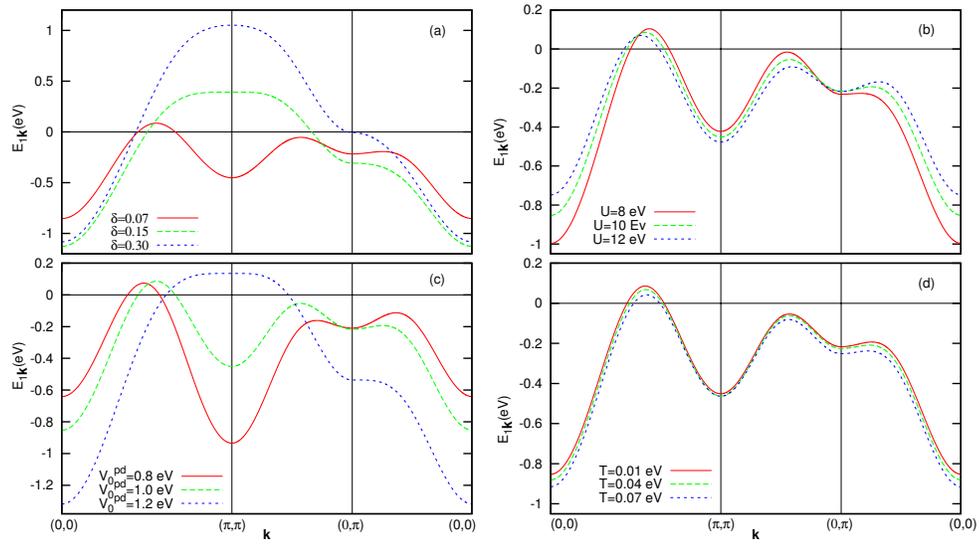}
\caption{\label{fig2} The quasiparticle bands intercepted by the chemical potential ($\mu=0$). In (a), the bands for  $U=10$ eV, $V_0^{pd}=1.0$ eV, $k_BT=0.01$ eV and three different doping levels $\delta$. In (b), the bands for  $\delta=0.07$, $V_0^{pd}=1.0$ eV, $T=0.01$ eV and three distinct intensities of Coulomb interaction $U$. In (c), the bands for  $\delta=0.07$, $U=10$ eV, $k_BT=0.01$ eV and three different hybridizations. The figure (d) shows the quasiparticle bands for $\delta=0.07$, $U=10$ eV,  $V_0^{pd}=1.0$ eV and three different temperatures. }
\end{figure}
Figure \ref{fig2}a displays the quasiparticle band for distinct doping levels $\delta$. While in the over and moderated doped regimes the quasiparticle bands cross the Fermi level near $(\frac{\pi}{2},\frac{\pi}{2})$ and the antinodal point $(0,\pi)$, in the underdoped regime the quasiparticle band crosses the Fermi level twice nearer the nodal point $(\frac{\pi}{2},\frac{\pi}{2})$. Such a behavior gives rise to a pocket around $(\frac{\pi}{2},\frac{\pi}{2})$ (see also figure \ref{fig1}c). On the other hand, as the quasiparticle band does not touch the Fermi level near $(0,\pi)$, a pseudogap emerges at that region. The kink observed near the $(\pi,\pi)$ point of the quasiparticle band is caused by
the strong antiferromagnetic correlations associated with $\langle \vec{S_j}\cdot\vec{S_i}\rangle$, which are maximum in {\bf Q}$=(\pi,\pi)$. The {\bf Q} is the antiferromagnetic wave-vector.

\begin{figure}
 \centering
\includegraphics[angle=-90,width=12 cm]{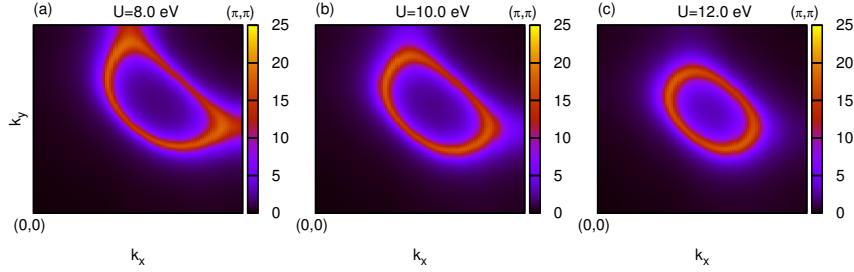}
\caption{\label{fig3} 
The  spectral function $A(\vec{k},\omega = 0)$ for 
$\delta=0.07$ and $V_0^{pd}=1.0$ eV and different intensities of Coulomb interaction. The remaining model parameters are identical to figure \ref{fig1}.}
\end{figure}

\begin{figure}
 \centering
\includegraphics[angle=-90,width=12 cm]{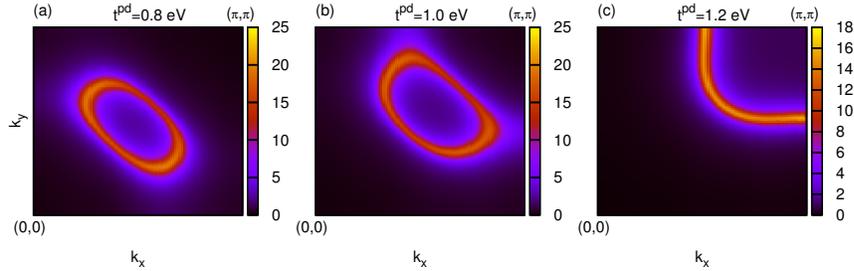}
\caption{\label{fig4}The spectral function $A(\vec{k},\omega = 0)$ representing the Fermi surface for different hybridizations $V_{0}^{pd}=t^{pd}$. The model parameters considered here are $U=10$ eV, $t^d=-0.5$ eV, $t^{\ell d}=0.1$ eV and $k_BT=0.01$ eV.}
\end{figure}

Figures \ref{fig2}b, \ref{fig2}c and \ref{fig2}d show the lower quasiparticle band for different $U$, $V_0^{pd}$ and $T$, respectively. In \ref{fig2}b and \ref{fig2}c, it can be noted that while the Coulomb interaction $U$ moves the quasiparticle band near $(\frac{\pi}{2},\pi)$ to lower energies, the hybridization $V_0^{pd}$ moves it to greater energies. Indeed, $U$ and $V_0^{pd}$ produce opposite effects on the pseudogap, i.e., $U$ increases the width of the pseudogap while $V_0^{pd}$ suppresses the pseudogap. Such a behavior of the pseudogap is also clear observed in figures \ref{fig3} and \ref{fig4}.
\begin{figure}
 \centering
\includegraphics[angle=-90,width=13cm]{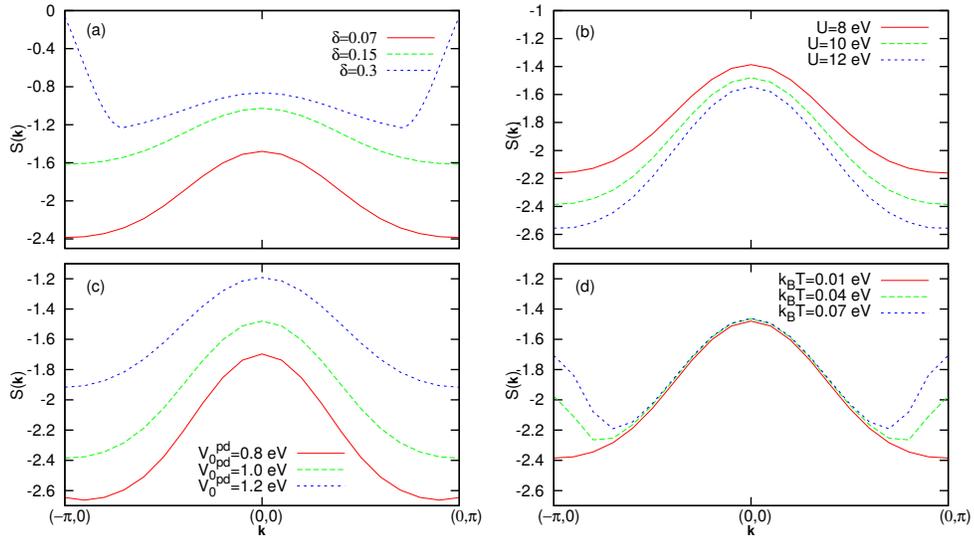}
\caption{\label{fig5}  (a) The Fourier transform $(S(\vec{k}))$ of the spin-spin correlation function $\langle \vec{S}_j\cdot\vec{S}_i\rangle$ for $U=10$ eV, $V_0^{dp}=1.0$ eV, $k_BT=0.01$ eV and different doping levels. (b) The $S(\vec{k})$ for $\delta= 0.07$ and different intensities of Coulomb interaction. The remaining parameters are identical to (a).
(c) The behavior of $S(\vec{k})$ for $\delta= 0.07$, $U=10$ eV, $k_BT=0.01$ eV and distinct values of $d-p$ hybridization. (d) The effect of temperature on $S(\vec{k})$ for $\delta= 0.07$, $U=10$ eV and $V_0^{dp}=1.0$ eV. The model parameters considered here are $t^d=-0.5$ eV and $t^{\ell d}=0.1$ eV.}
\end{figure}

The features observed in the Fermi surfaces presented in figures \ref{fig1}, \ref{fig3} and \ref{fig4} are understood in terms of the antiferromagnetic correlations associated with the spin-spin correlation function 
$\langle \vec{S}_j\cdot\vec{S}_i\rangle$ introduced in equations (\ref{h2}) and (\ref{SjSi2}). Actually, $S(\vec{k})$ which is the Fourier transform of  $\langle \vec{S}_j\cdot\vec{S}_i\rangle$, is analyzed in the $\vec{k}$-space. Figure \ref{fig5} shows $S(\vec{k})$ at the directions $(-\pi,0)$-$(0,0)$-$(\pi,0)$ in the first Brillouin zone. In  \ref{fig5}(a), with different doping levels $\delta$, it is clear that $S(\vec{k})$ is high doping dependent, mainly in the underdoped region close to the antiferromagnetic insulator (AFI) phase.  For $\delta=0.30$, the chemical potential $\mu$ 
is moved to lower energies due to the low occupation $n_T$. As a consequence, the Fermi function removes states above the chemical potential leading to a suppression of $S(\vec{k})$ for $\vec{k}>\vec{k}_F$ ($\vec{k}_F$ is the Fermi wave vector).
The doping region, from where the pseudogap emerges, coincides with that in which $S(\vec{k})$ is very strong.
In figure \ref{fig5}(b), the effect of the Coulomb interaction $U$ on $S(\vec{k})$ is investigated. As it can be verified, $U$ increases the intensity of $S(\vec{k})$ mainly at the antinodal points region.
On the other hand, it has been verified that the $d-p$ hybridization $V_0^{dp}$ acts on $S(\vec{k})$ decreasing its intensity (see figure \ref{fig5}(c)). Finally, figure \ref{fig5}(d) shows $S(\vec{k})$ for different temperatures. It is interesting to notice that the temperature decreases $S(\vec{k})$ especially in the region of the antinodal points where the pseudogaps are present. This behavior comes from the effect of the temperature on the Fermi function, i.e., the Fermi function increases it slope near the Fermi level, suppressing $S(\vec{k})$ when the temperature is enhanced. If the results from figures \ref{fig5}(a), \ref{fig5}(b) and \ref{fig5}(c)  are compared with those ones from figures  \ref{fig1}, \ref{fig3} and \ref{fig4}, it is clear that the presence of pseudogaps on the Fermi surface is directly related to the intensity of $S(\vec{k})$ and, therefore, to antiferromagnetic correlations. 

It is worth to notice that in the overdoped regime, where the correlations associated with $S(\vec{k})$ are weakened, the effects of the hybridization are not very significant. In the underdoped regime, where mostly sites are single occupied, the hopping processes require a double occupied site. 
As a consequence, the electrons tend to stay localized due to the high cost of energy ($U$) to double occupy a site. 
Nevertheless, if the hybridization is present, the hopping to a $p$ orbital in a site $j$, via hybridization, can occur. Thus, if the hybridization is favored, the occupation of the $d$-band may decrease depleting the spin-spin correlations. This scenario  allows understanding the effects of the hybridization on the antiferromagnetic correlations (associated with  $\langle \vec{S}_j\cdot\vec{S}_i\rangle$) and, consequently, on the Fermi surface topology.

\section{Conclusions}

 In the present work, the original two-poles approximation proposed by Roth \cite{Roth69} has been improved in order to consider the correct momentum dependence of the spin-spin correlation function $\langle \vec{S}_j\cdot\vec{S}_i\rangle$. The structure of $\langle \vec{S}_j\cdot\vec{S}_i\rangle$ in the momentum space is essential to catch important effects due to antiferromagnetic corrrelations which, in the present approach, are the source of anomalies as pseudogap and hole-pockets on the Fermi surface. The role of the antiferromagnetic correlations associated with the spin-spin correlation function has been investigated in different situations.
Initially, it is shown that the 
hole underdoped regime is characterized by the presence of hole-pockets enclosing the nodal points and pseudogaps near the antinodal points. The results show also that the Coulomb interaction increases the region (in the $\vec{k}$ space) of the pseudogap and decreases the area enclosed by the hole-pockets.
On the other hand, the $V_0^{dp}$ hybridization acts in the sense of to decreasing the region (in the $\vec{k}$ space) where the pseudogap occurs. If the $V_0^{dp}$ is sufficiently high, the pseudogap vanishes and an ordinary large Fermi surface is obtained. The scenario for the Fermi surface described above, can be understood in terms of the antiferromagnetic correlations related to the spin-spin correlation function $\langle \vec{S}_j\cdot\vec{S}_i\rangle$.
The hole doping $\delta$  damps down the intensity of the correlation function $\langle \vec{S}_j\cdot\vec{S}_i\rangle$. From the overdoped regime up to moderated doping levels $(\delta ~ 0.1 )$, the antiferromagnetic correlations are weak and the Fermi surfaces are typical of a normal metal without anomalies. In summary, it has been shown that in the underdoped regime, where $\langle \vec{S}_j\cdot\vec{S}_i\rangle$ is very strong, the pseudogap and the hole-pockets emerge. The local Coulomb interaction $U$ increases the $\langle \vec{S}_j\cdot\vec{S}_i\rangle$ favoring the pseudogap and the hole-pockets. On the other hand, the $V_0^{dp}$ hybridization suppresses the $\langle \vec{S}_j\cdot\vec{S}_i\rangle$ recovering a large Fermi surface as those ones observed in the overdoped regime. Finally, the temperature acts on $\langle \vec{S}_j\cdot\vec{S}_i\rangle$ decreasing it, mainly, at the antinodal points which are the region of the pseudogap in the $\vec{k}$ space.

To conclude, this work has presented a description for the hole-doped regime of the $d-p$ Hubbard model within the n-pole approximation. Par\-ti\-cularly, in the underdoped regime, it has been shown a route which leads to pockets and pseudogap in the Fermi surface.
It should be remarked that the correlation function $\langle \vec{S}_j\cdot\vec{S}_i\rangle$ present in the band shift $W_{\vec{k}\sigma}^{d}$ plays an important role in the underdoped situation. More precisely, antiferromagnetic correlations coming from $\langle \vec{S}_j\cdot\vec{S}_i\rangle$
strongly modify the quasi-particle band structure.  This is the ultimate source of anomalies in the Fermi surface in the present approach.

\section*{Acknowledgments}
This work was partially supported by the Brazilian agencies CNPq (Conselho Nacional de Desenvolvimento Cient\'{\i}fico e Tecnol\'ogico), CAPES (Coordena\c{c}\~ao de Aperfei\c{c}oamento de Pessoal de N\'{\i}vel Superior) and FAPERGS (Funda\c{c}\~ao
de Amparo \`a Pesquisa do Rio Grande do Sul).

\end{document}